\documentclass[a4paper,twocolumn,showpacs,amsmath, amssymb,nofootinbib,aps,floatfix]{revtex4}
\usepackage{epsfig}
\input psfig.sty 

\usepackage{bm}									
\usepackage{dcolumn}								
\usepackage{graphicx}								
\usepackage{hyperref}

\newcommand{\no}{\mbox{$n_{e_o}$}}

\newcommand{\dTo}{\mbox{$\Delta T_0$}}
\newcommand{\Xo}{\mbox{$S_{x  0}$}}

\newcommand{\LameH}{\mbox{$\Lambda_{e \mbox{\tiny H}}$}}

\newcommand{\sigT}{\mbox{$\sigma_{\mbox{\tiny T}}$}}
\newcommand{\Tcmb}{\mbox{$T_{\mbox{\tiny CMB}}$}}

\newcommand{\nH}{\mbox{$n_{\mbox{\tiny H}}$}}

\newcommand{\Da}{\mbox{$D_{\!\mbox{\tiny A}}$}}

\newcommand{\Mtot}{\mbox{$M_{\mbox{\scriptsize total}}$}}

\begin{document}

\title{A test for cosmic distance duality}

\author{R. F. L. Holanda$^{1,2}$\footnote{E-mail: holanda@on.br}}

\author{R. S. Gon\c{c}alves$^1$\footnote{E-mail: rsousa@on.br}}

\author{J. S. Alcaniz$^1$\footnote{E-mail: alcaniz@on.br}}

\address{$^1$Departamento de Astronomia, Observat\'orio Nacional, 20921-400, Rio de Janeiro - RJ, Brasil}

\address{$^2$Departamento de F\'{\i}sica, Universidade Estadual da Para\'{\i}ba, 58429-500, Campina Grande - PB, Brasil}

\date{\today}

\begin{abstract}
Testing the cosmic distance duality relation (CDDR) constitutes an important task for cosmology and fundamental physics since any violation of it would be a clear evidence of new physics. In this {paper}, we propose a new test for the CDDR using only current measurements of the gas mass fraction of galaxy clusters from Sunyaev-Zeldovich  ($f_{SZE}$) and X-ray surface brightness ($f_{X-ray}$) observations. We show that the relation between  $f_{X-ray}$ and $f_{SZE}$ observations is given by $f_{SZE}=\eta f_{X-ray}$, where $\eta$ quantifies deviations from the CDDR. Since this latter expression is valid for the same object in a given galaxy cluster sample, the method proposed removes possible contaminations from different systematics error sources and redshift differences involved in luminosity and angular diameter distance measurements. We apply this cosmological model-independent methodology to the {most recent $f_{X-ray}$ and $f_{SZE}$ data} and show that no significant violation of the CDDR is found.

\end{abstract}

\pacs{98.80.-k, 98.80.Es, 98.65.Cw}

\maketitle

\section{Introduction}

The so-called cosmic distance duality relation (CDDR)~\cite{ellis71}, which is closely connected with the Etherington reciprocity theorem~\cite{tolman, Etherington33}, plays an important role in observational cosmology ranging from gravitational lensing studies, galaxy and galaxy clusters observations, to analyses of the cosmic microwave blackbody radiation (CMB)\cite{SEF99}. It relates the {luminosity distance} $D_{\scriptstyle L}$ with the {angular diameter distance} $D_{\scriptstyle A}$ through the identity
\begin{equation}
  \eta=\frac{D_{\scriptstyle L}}{D_{\scriptstyle A}}{(1+z)}^{-2}=1\;.
  \label{rec3}
\end{equation}

 The above result can be easily demonstrated for usual Friedmann-Lema\^itre-Robertson-Walker cosmologies. As is well known, it only requires source and observer to be connected by null geodesics in a Riemannian spacetime and cosmological conservation of the number of photons~\cite{Etherington33,ellis71}. Examples of non-standard scenarios that violate the equality (\ref{rec3}) are models in which photons do not travel on unique null geodesics, models with variations of fundamental constants or with photon non-conservation due to coupling to particles beyond the standard model of particle physics, absorption by dust, etc. (see, e.g., \cite{Csaki02,Jaeckel10,lverde} for a discussion).

 Attempts to actually test the CDDR validity from astronomical observations have started only recently \cite{Uzan04,DeBernardis,Holanda10}. For instance, in Ref.~\cite{Uzan04} it was argued that the Sunyaev-Zel'dovich effect (SZE) plus X-ray  technique for measuring galaxy cluster distances is strongly dependent on the CDDR validity. As is well known, one can consider different electronic density dependencies combined with some assumptions about the galaxy cluster morphology in order to evaluate its angular diameter distance with basis on Eq.\ (\ref{rec3}), i.e. \cite{Carlstrom2002}
\begin{equation}
D_{A}^{data}(z)\propto \frac{(\Delta
T_{0})^{2}\Lambda_{eH0}}{(1+z)^4
S_{X0}{T_{e0}}^{2}}\frac{1}{\theta_{c}},
\end{equation}
where $D_{A}^{data}(z)$ is the angular diameter distance to a given galaxy cluster obtained directly from a jointly analysis of its SZE and X-ray surface brightness observations, $S_{X0}$ is the central X-ray surface brightness, $T_{e0}$ is the central temperature of the intra-cluster medium, $\Lambda_{eH0}$ is the central X-ray cooling function of the intra-cluster medium, $\Delta T_0$ is the central decrement temperature, and $\theta_{c}$ refers to a characteristic scale of the cluster along the line of sight (l.o.s.), whose exact meaning depends on the assumptions adopted to describe the galaxy cluster morphology.

However, if one assumes a more general expression for CDDR, such that,
\begin{equation}
\frac{D_{\scriptstyle L}}{D_{\scriptstyle A}}{(1+z)}^{-2}= \eta (z),
\label{receta}
\end{equation}
it is possible to show  that (for details see \cite{Uzan04,Holanda11})
\begin{equation}
D_{A}^{\: data}(z)=D_{A}(z)\eta(z)^{2}. \label{recc}
\end{equation}
Therefore,  $D_{A}^{data}(z)$  reduces to the real angular diameter distance  only when the DD relation is strictly valid ($\eta\equiv 1$). In order to quantify the $\eta$ parameter, the authors in Ref.\cite{Uzan04}
fixed $D_A(z)$ by using the cosmic concordance model \cite{spergel03} while for $D^{\: data}_{A}(z)$ they considered the 18 galaxy clusters from the Ref.\cite{Reese02} for which a spherically symmetric cluster geometry has been assumed. By assuming $\eta$ constant, their statistical analysis provided $\eta = 0.91^{+0.04}_{-0.04}$ (1$\sigma$), and is therefore only marginally consistent with the standard result.

On the other hand,  the CDDR should be tested only from astronomical observations, i.e., finding cosmological sources whose intrinsic luminosities and  intrinsic sizes are known. Thus,  after measuring the source redshift, one can determine both $D_{\scriptstyle L}$ and $D_{\scriptstyle A}$ to test directly the relation. In principle, this ideal method should not use any relationship coming from a specific cosmological model, i.e., they must be determined by means of intrinsic astrophysical quantities only.

In recent papers, the validity of the {CDDR} has been discussed using $D_A$ measurements from galaxy clusters (GC) and luminosity distances from type Ia supernovae (SNe Ia) \cite{Holanda10}. In such analyses,  subsamples of GC and SNe Ia are built so that differences in redshift between objects in each sample are small ($\Delta z \simeq 10^{-3}$), thereby allowing a validity test of the CDDR. There are, however, at least three aspects that should be considered when performing this kind of analysis. First, that estimates of $\eta$ from this method are potentially contaminated by different systematics error sources in GC and SNe Ia observations. Second, that some SNe Ia light curve fitters use a specific cosmological scenario in their calibration process, which makes the GC/SNe Ia test not completely model-independent (see Ref. \cite{Holanda11} for a discussion on  influence of the SNe Ia light curve fitters on the CDDR test). Finally, that the use of different objects to derive $D_{\scriptstyle L}$ and $D_{\scriptstyle A}$ implies necessarily in a choice for $\Delta z$, which affects the resulting estimates of the $\eta$ parameter (see \cite{shu} for a recent discussion. See also~\cite{Khedekar11} for an interesting model-independent CDDR test  based on future observations of a redshifted 21cm signal from disk galaxies).

In order to circumvent these observational problems, we propose a consistent model-independent test for Eq. (\ref{rec3}) that uses only {current observations of  the gas mass fraction of GC's}. To perform our analysis, we use {a  sample of $f_{X-ray}$ and $f_{SZE}$ measurements of 38 GC's as discussed in Ref.~\cite{LaRoque06}.  We show that when the CDDR is taken into account current  $f_{X-ray}$ and $f_{SZE}$ measurements provide a direct test for the CDDR.}

\section{Gas mass fraction}

\subsection{X-ray Observations}

The gas mass fraction is defined as $f =M_{gas}/M_{Tot}$ \cite{Sasaki96}, where $M_{Tot}$ is the total mass obtained via hydrostatic equilibrium assumption and  $M_{gas}$ (gas mass) is obtained by  integrating the gas density model. The gas density model frequently used is the 3-dimensional electron number density of the spherical $\beta$ model given by \cite{Cavaliere78}\footnote{It is worth mentioning that we have used $f_{gas}$ data for the non-isothermal double $\beta$-model in our statistical analyses. The spherical $\beta$ model is used in this section only for simplicity but without loss of generality for the test proposed.}
\begin{equation}
n_e({\mathbf{r}}) = \no \left (1 + \frac{r^2}{r_c^2} \right )^{-3\beta/2},
\label{eq:single_beta}
\end{equation}
where $n_e$ is the electron number density, $r$ is the radius from the center of the cluster, $r_c$ is the core radius of the intracluster medium (ICM), and $\beta$ is a power law index. Under this assumption, $M_{tot}$ and $M_{gas}$ are given by \cite{Grego01}
\begin{equation}
\Mtot(r)=\frac{3\beta k_{\rm B} T_{\rm e}}{G \mu m_p} \frac{r^3}{r_c^2 + r^2},
\label{eq:mtot_hse_iso}
\end{equation}
 where $T_{\rm e}$ is the temperature of the intra cluster medium, $\mu$ and $m_p$ are, respectively, the total mean molecular weight and the proton mass, $k_{\rm B}$ the Boltzmann constant, and
\begin{equation}
M_{gas}(r) = A \int_{0}^{r/\Da} \left (1+\frac{\theta^2}{\theta_c^2}
\right)^{-3\beta/2}\: \theta^2 d\theta,
\label{eq:mgas_single}
\end{equation}
where $A=4\pi \mu_e \no m_p\, \Da^3$, $\theta=r/D_A$ and $\mu_e$ is the mean molecular weight of the electrons.

The X-ray surface brightness over some frequency band is written as
\begin{equation}
S_x = \frac{D_A^{2}}{4\pi D^2_L} \! \int \!\!  \, n_e \nH \LameH \,d\ell
\label{eq:xray_sb}
\end{equation}
where the integral is along the line of sight and $\LameH$ is the X-ray cooling function, proportional to $T_e^{1/2}$ ~\cite{Sarazin88}.  Note that the above equation depends explicitly on the ratio between the angular diameter and luminosity distances. Thus, if the CDDR is taken as valid [$\eta=1$ in Eq. (\ref{rec3})] and a fiducial angular diameter distance, $D_A^{*}$, is assumed,  the central electron density $n_{e0}$ can be  analytically obtained, i.e.,  \cite{LaRoque06}
\begin{equation}
n^{X-ray}_{e0} = \left( \frac{\Xo \:4\pi (1+z)^4 \:\frac{\mu_H}{\mu_e}\:
  \Gamma(3\beta)}{\LameH D_A^{*} \pi^{1/2}\: \Gamma(3\beta-\frac{1}{2})\,
  \theta_c} \right)^{1/2},
\label{eq:xray_ne0}
\end{equation}
{providing the well-known relation in current  X-ray gas mass fraction measurements $f_{X-ray} \propto {D_A^*}^{3/2}$~\cite{Sasaki96} (in the above expression $S_{x0}$ is the central surface brightness and $\Gamma(x)$ is the Gamma function).}

However, if the validity of the CDDR is not previously assumed ($\eta \neq 1$),  Eq. (\ref{eq:xray_ne0}) is now rewritten as
\begin{equation}
n^{X-ray}_{e0} = \eta \left( \frac{\Xo \:4\pi (1+z)^4 \:\frac{\mu_H}{\mu_e}\:
  \Gamma(3\beta)}{\LameH \Da \pi^{1/2}\: \Gamma(3\beta-\frac{1}{2})\,
  \theta_c} \right)^{1/2}\;,
\label{eq:xray_ne02}
\end{equation}
which clearly shows that gas mass fraction measurements  extracted from X-ray data are affected by a possible violation of the CDDR scaling as
\begin{equation}
f_{X-ray}^{th} \propto \eta {D_A}^{3/2}\;.
\end{equation}

\begin{figure*}
{\includegraphics[width=58mm, height=62mm, angle=0]{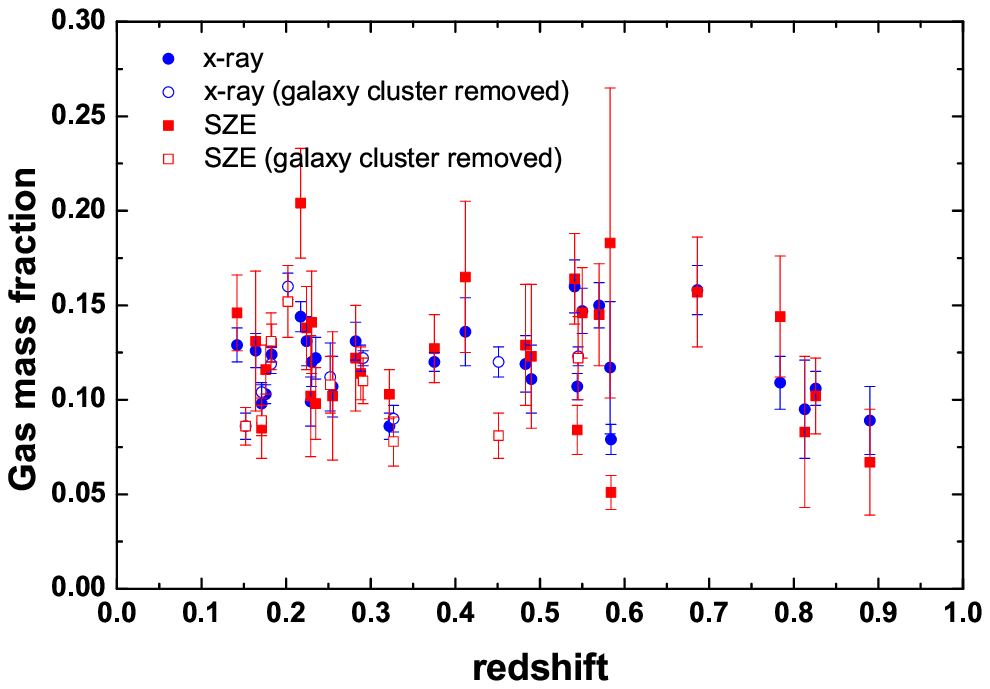}
\includegraphics[width=58mm, height=62mm, angle=0]{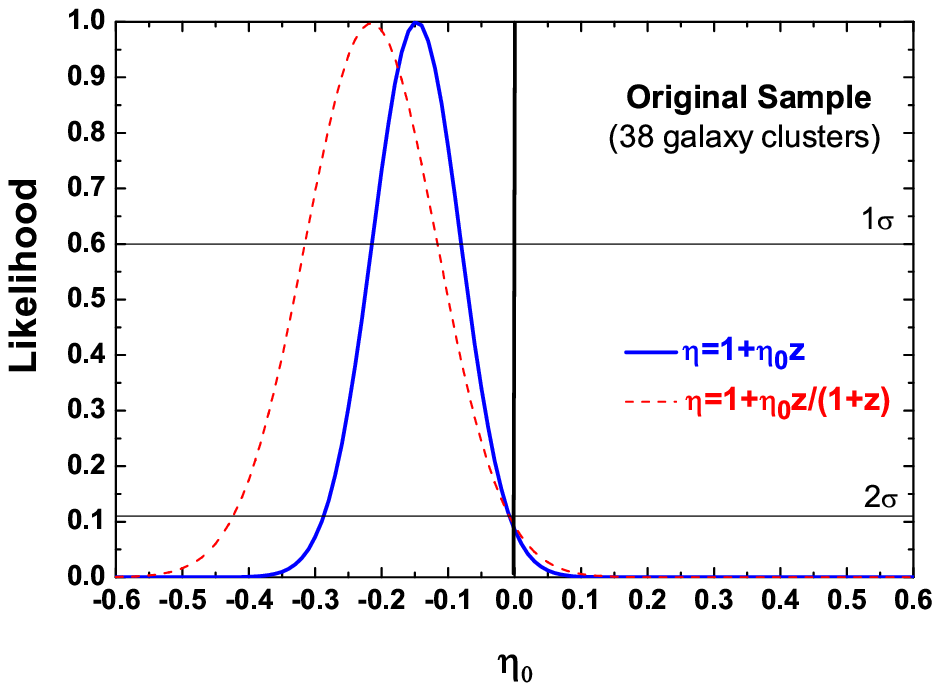}
\includegraphics[width=58mm, height=62mm, angle=0]{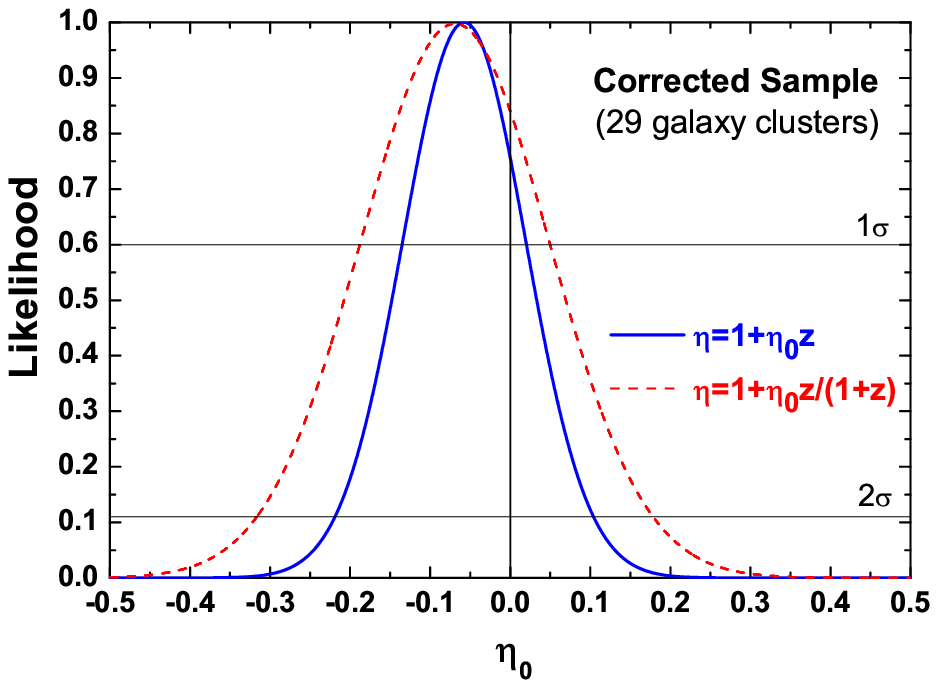}
\hskip 0.1in} \caption{{\bf{a)}} Gas mass fraction measurements as a function of redshift for 38 galaxy clusters~\cite{LaRoque06}. Blue circles (red squares) stand for measurements obtained via current X-ray (SZE) observations. Open circles and squares correspond to galaxy clusters that present large reduced $\chi^{2}$ ($\chi_{d.o.f.}^2  > 2$) when described by the hydrostatic equilibrium model. {\bf{b)}} The likelihood distribution functions for the CDDR parameter $\eta_0$. Blue solid lines correspond to the results for P1 whereas red dashed lines for P2. In this analysis, the entire sample of Ref.~\cite{LaRoque06} was used (38 clusters). {\bf{c)}} The same as in Panel 1b for the reduced galaxy cluster sample (29 clusters). When compared with the previous analysis, the compatibility of the data with the CDDR is clearly more evident.} \label{fig:Analysis}
\end{figure*}

\subsection{Sunyaev-Zel'dovich Observations}\label{sze}

The measured temperature decrement $\Delta T_{\rm SZE}$ of the CMB due to the Sunyaev-Zel'dovich effect \cite{SunZel72} is given by \cite{LaRoque06}
\begin{equation}
\label{eq:sze1} \frac{\Delta T_{\rm 0}}{T_{\rm CMB}} = f(\nu,
T_{\rm e}) \frac{ \sigma_{\rm T} k_{\rm B} }{m_{\rm e} c^2} \int n_e T_{\rm e} dl, \
\end{equation}
where $T_{\rm CMB} =2.728$ K
is the present-day temperature of the CMB, $\sigma_{\rm T}$ the Thompson cross
section, $m_{\rm e}$ the electron mass and $f(\nu, T_{\rm e})$
accounts for frequency shift and relativistic corrections~\cite{itoh98}. It worth mentioning that the gas temperature, $T_e$, is  insensitive to the validity of the CDDR since it is obtained through the shape of the  X-ray spectra (bremsstrahlung spectra) and not through  X-ray luminosity.

Using SZE observations, the central electron density can now be expressed as

\begin{equation}
n_{e0}^{SZE} = \left( \frac{\dTo \,m_e c^2 \:\Gamma(\frac{3}{2}\beta)}{f_{(\nu, T_e)}
  \Tcmb \sigT\, k_{\rm B} T_e \Da \pi^{1/2} \:\Gamma(\frac{3}{2}\beta -
  \frac{1}{2})\, \theta_c} \right),
\label{eq:sz_ne0}
\end{equation}
which is insensitive to the validity of the CDDR. Therefore, current gas mass fraction measurements via SZE  depend only on the  angular diameter distance as
\begin{equation}
f_{SZE} \propto {D_A}.
\end{equation}

\subsection{$f_{SZE}/f_{X-ray}$ relation}

Current $f_{X-ray}$ measurements have been obtained by assuming the validity of the CDDR~\cite{Ettori,Allen}. If, however, this is not the case, the real gas mass fraction from X-ray observations (Eq.(11)) should be related with the current  observations by $f_{X-ray}^{th}=\eta f_{X-ray}$. So, if all the physics behind the X-ray and SZE observations are properly taken into account, one would expect $f_{gas}$ measurements from both tecniques to agree with each other since they are measuring the very same physical quantity. In this way, it is clear that the general expression relating current X-ray  and  SZE  observations is given by:
\begin{equation} \label{relation}
f_{SZE}=f_{X-ray}^{th}=\eta f_{X-ray}\;,
\end{equation}
 which provides a direct test for the CDDR. Since the above expression holds for a given object, a possible influence on the $\eta$ estimates due to redshift differences of distinct objects  (e.g., in tests involving SNe Ia and GC) is fully removed.

Before discussing our estimates on the $\eta$ parameter from the above relation, it is worth mentioning that in both methods to test the CDDR (the one of Eq. (\ref{relation}) and those using $D_A$ measurements of galaxy clusters from their SZE and X-ray observations~\cite{Uzan04,Holanda10,Holanda11}) other physical effects could be present in the SZE observations, leading to results with $\eta \neq 1$ that not necessarily would be related to a CDDR  violation in X-ray frequency. For instance, as mentioned earlier  [see Eq. (13)], SZE observations are independent of the CDDR validity. However, these observations are redshift-independent  only if there is no process of energy injection into the CMB. Otherwise,  the standard linear relation of the CMB temperature evolution, $T_{CMB}(z)=T_{CMB}(z=0)(1+z)$, is not valid~\cite{Lima1996} and changes are needed in Eqs. (12) and (13). In this line, there are recent analyses testing the standard evolution of the CMB temperature through different techniques  by using a more general expression such as  $T_{CMB}(z)=T_{CMB}(z=0)(1+z)^{1+ \beta}$ \cite{Lima2000,Battistelli2002,Hole,Avgoustidies,Noterdaeme}. As a basic result, all analyses confirm  the standard relation, i.e., $\beta \approx 0$ (see Ref. \cite{Noterdaeme} for  recent constraints on $\beta$ value by using SZE observations and carbon monoxide excitation at high-$z$).

 In what follows, we explore Eq. (\ref{relation}) and discuss constraints on the CDDR from current X-ray and SZE gas mass fraction  measurements.

\section{Galaxy cluster Data}

To discuss the potential of  Eq. (\ref{relation}) in probing the CDDR, we use the most recent X-ray and SZE gas mass fraction  measurements to date, as given in Ref.~\cite{LaRoque06}. The  sample consists of  38 massive galaxy clusters spanning redshifts from 0.14 up to 0.89 whose X-ray data were obtained from the {Chandra X-ray Observatory} and SZE data from the BIMA/OVRO SZE imaging project, which uses the Berkeley-Illinois-Maryland Association (BIMA) and Owens Valley Radio Observatory (OVRO) interferometers to image the SZE. In order to perform a realistic model for the cluster gas distribution and take into account a possible presence of  cooling flow, the gas density was modeled  with the non-isothermal double $\beta$-model that generalizes the single $\beta$-model profile given by Eq. (\ref{eq:single_beta}).

Another important aspect worth mentioning  is that the shape parameters of the gas density model ($\theta_c$ and $\beta$) were obtained  from a joint analysis of the X-ray and SZE data~\cite{Bonamente04}, which makes the  SZE gas mass fraction not independent. However,  as has been shown from current simulations~\cite{hallman}, the values of $\theta_c$ and $\beta$ computed separately by SZE and X-ray observations agrees at $1\sigma$ level within a radius $r_{2500}$ (at which the mean enclosed mass density is equal to 2500 cosmological critical density), the same  used in the La Roque {\it{et al.}} observations. Therefore, we believe that, unless a violation of the CDDR may affect even the shape parameters obtained  from the X-ray data (and not only $S_{x0}$), one can use the  La Roque {\it{et al.}} sample to perform the method discussed earlier. In Figure 1a we show the sample of 38 measurements of the gas mass fraction of galaxy clusters obtained via X-ray surface brightness ({ assuming $\eta=1$) and the SZE (see also Table 5 of Ref. \cite{LaRoque06})}.


\section{Analysis and Results}

We modify the {CDDR} to test for any violation by using simple parameterizations for $\eta$. In order to take into account a possible influence of different $\eta$ parameterizations on the results, we use in our analyses two functions~\cite{Holandaaa}:
\begin{eqnarray}
\label{parameterizations}
\eta(z) = \; \left\{
\begin{tabular}{l}
$1 + \eta_{0} z$
\,
\quad \quad  \quad  \quad \quad \quad \quad \quad (P1)\\
\\
$ 1 + \eta_{0}z/(1+z)$
\quad \quad  \quad \quad \quad (P2) \\
\end{tabular}
\right.
\nonumber
\end{eqnarray}
P1 is a continuous and smooth one-parameter linear expansion, whereas P2 includes a possible epoch-dependent correction, which avoids the divergence at very high $z$. Note also that, differently from constant $\eta$ parameterizations (e.g., $D_L/D_A = \eta_0(1+z)^{2}$~\cite{DeBernardis}), both P1 and P2 recover the CDDR in the limit $z \rightarrow 0$, as obtained from cosmographic derivations.

We evaluate our statistical analysis by defining the likelihood distribution function ${\cal{L}} \propto e^{-\chi^{2}/2}$, where
\begin{equation}
\label{chi2} \chi^{2} = \sum_{i = 1}^{N}\frac{{\left[\eta(z) - \eta_{i, obs}(z) \right] }^{2}}{\sigma^{2}_{i, obs}},
\end{equation}
{$\eta_{i, obs}(z) = f_{SZE}/f_{X-ray}$ and $\sigma^{2}_{i, obs}$ is the uncertainty }associated to this quantity. Also, the $i$ index indicates the sum over each cluster. In Fig. 1b  we plot our first constraints on the {CDDR} from Eq. (\ref{relation}). By considering all 38 galaxy clusters of the sample discussed earlier, we show the likelihood distribution as a function of the parameter $\eta_0$ for P1 (blue solid line) and P2 (red dashed line). For these two cases, we find $\eta_{0} = -0.15 \pm {0.07}$ ($\chi_{d.o.f.}^2 = 1.02$) and  $\eta_{0} = -0.22 \pm {0.10}$ ( $\chi_{d.o.f.}^2 = 1.04$)  at 1$\sigma$ level, respectively.  Note that the {CDDR} ($\eta_0 = 0$) is slightly compatible with theses data, being $\simeq 2\sigma$ off from the best-fit values in both cases. For the sake of completeness and also to verify the effect of the gas modelling on the results, we also performed the same analysis using the isothermal $\beta$ model. In this case, we obtained $\eta_0 = -0.08 \pm 0.16$ and $\eta_0 = -0.13 \pm 0.21$ at $2\sigma$ level for the linear and non-linear parametrizations, respectively.

An important aspect concerning the galaxy cluster sample shown in Fig. 1a is that some objects present questionable  reduced $\chi^{2}$ ($2.43 \leq \chi_{d.o.f.}^2 \leq 41.62$) when described by the hydrostatic equilibrium model (see Table 6 in Ref. \cite{Bonamente06}). They are: Abell 665, ZW 3146, RX J1347.5-1145, MS 1358.4 + 6245, Abell 1835, MACS J1423+2404, Abell 1914, Abell 2163, Abell 2204. By excluding these objects from our sample (we end up with a subsample of 29 galaxy clusters), we perform a new analysis whose results are displayed in Fig. 1c. We note that, when compared with the previous analysis (Fig. 1b), the compatibility of the data with the CDDR is clearly more evident now, with $\eta_{0} = -0.06 \pm {0.07}$ (P1) and $\eta_{0} = -0.07 \pm {0.12}$ (P2)  at 1$\sigma$ level. It is worth observing that for all analyses performed in this paper a negative value for the CDDR parameter was prefered by the data (although the data are fully consistent with $\eta_0 = 0$). A possible explanation for that has been discussed in Ref.~\cite{lverde} in terms of cosmic opacity or the existence of axion-like and mini-charged particles (see, e.g., \cite{Jaeckel10} for a recent review on these weakly-interacting-sub-eV particles). In Table I we summarize the main results of our analyses.

\begin{table}
\begin{center}
\begin{tabular} {|c|c|c|}
\hline\hline  38 galaxy clusters data& $\chi^{2}/d.o.f$
\\ \hline \hline
P1 \quad \quad $\eta_{0} =  -0.15\pm  0.14$ & 1.02    \\
P2 \quad \quad $\eta_{0} = -0.22\pm 0.21$ &1.04 \\
\hline\hline  29 galaxy clusters data& $\chi^{2}/d.o.f$
\\ \hline \hline
P1 \quad \quad $\eta_{0} = -0.06\pm 0.16$  & 0.93    \\
P2 \quad \quad $\eta_{0}  = -0.07 \pm 0.24$ & 0.92  \\
\hline \hline
\end{tabular} \label{table1}
\caption{Constraints on the CDDR parameter $\eta_0$. The error bars correspond to 2$\sigma$ ($\Delta \chi^2 = 4$).}
\end{center}
\end{table}

\section{conclusions}

In this {\it Letter}, we have discussed how {current measurements} of  gas mass fraction of galaxy clusters from X-ray and SZE effect observations can be used to test the CDDR. We have shown that if this relation is consistently taken into account current $f_{X-ray}$ and $f_{SZE}$ measurements are related by $f_{SZE}=\eta f_{X-ray}$ [Eq. (\ref{relation})], allowing a direct test for the CDDR.

To perform our analyses we have considered two distinct forms for $\eta(z)$, i.e., $\eta = 1+\eta_{0}z$ and $\eta = 1+\eta_{0}z/(1+z)$, which recover the equality between $D_L$ and $D_A$ at very low redshifts. By considering 38 gas mass fraction measurements of galaxy clusters obtained from their SZE and X-ray emissions, we have found no significant influence of the above $\eta(z)$ parameterizations on the results and that a no violation of the CDDR is compatible at $2\sigma$ level. However, if our sample is corrected to account for possible statistical error sources due to galaxy cluster modeling, a value compatible with the validity of the CDDR is found (see Table I).

Finally, it is worth mentioning that,  irrespective of the fiducial model adopted in observations,  $f_{gas}$  measurements from both tecniques (${X-ray}$ and ${SZE}$) for a given cluster must agree  and if they do not the detected difference must be associated with the duality parameter $\eta$. Therefore, the  method here proposed is cosmological model-independent and also presents a clear advantage over tests involving different kinds of observations (e.g., GC and SNe Ia) since the above relation is analyzed for the same object in a given sample.  We believe that when applied to upcoming observational data the method discussed here may be useful to probe a possible violation of the CDDR.

\begin{acknowledgments}
The authors are very grateful to R. Dupke  and B. Santos for helpful discussions. JSA ans RGS thank CNPq and CAPES for the grants under which this work was carried out. RFLH acknowledges financial support from the Funda\c{c}\~ao de Amparo \`a Pesquisa do Estado do Rio de Janeiro (FAPERJ).
\end{acknowledgments}

\end{document}